# Phronesis of AI in radiology: Superhuman meets natural stupidity

**Judy W. Gichoya** [1], **Siddhartha Nuthakki** [2], **Pallavi G. Maity** [2], **Saptarshi Purkayastha** [2]
[1] Dept. of Radiology and Image Sciences, Indiana University School of Medicine;
[2] Dept. of BioHealth Informatics, Indiana University – Purdue University Indianapolis
jgichoya@iupui.edu, snuthakk@iu.edu, pmaity@iu.edu, saptpurk@iupui.edu

## Abstract

Advances in AI in the last decade have clearly made economists, politicians, journalists, and citizenry in general believe that the machines are coming to take human jobs. We review "superhuman" AI performance claims in radiology and then provide a self-reflection on our own work in the area in the form of a critical review, a tribute of sorts to McDermott's 1976 paper, asking the field for some self-discipline. Clearly there is an opportunity to replace humans, but there are better opportunities, as we have discovered to fit cognitive abilities of human and non-humans. We performed one of the first studies in radiology to see how human and AI performance can complement and improve each other's performance for detecting pneumonia in chest X-rays. We question if there is a practical wisdom or *phronesis* that we need to demonstrate in AI today as well as in our field. Using this, we articulate what AI as a field has already and probably can in the future learn from Psychology, Cognitive Science, Sociology and Science and Technology Studies.

## 1 Introduction

There is increasing application of AI, particularly computer vision using deep learning techniques to medical imaging datasets. Following the age-old trend of reporting AI performance benchmarked against humans in games, like the Chinook in Checkers [Schaeffer, 2007], IBM Deep Blue in Chess [Kelly III, 2015], IBM Watson in Jeopardy [Kelly III, 2015], AlphaGo in Go or Libratus in Poker [Brown and Sandholm, 2017], did we expect AI in medicine to be left far behind? In an evaluation of the performance of AI to diagnose three eye diseases, the authors report human level performance that is comparable to the current clinical model whose gold standard is based on assessment of retinal images by ophthalmologists. Improvements on model performance are again compared to human performance as exemplified by the updated version of retinal images deep learning that is now at the level of retinal subspecialists [Jonathan *et al*., 2018]. The Stanford AI group has also reported human level performance to detect pneumonia when compared to expert radiologists [Rajpurkar *et al*., 2017].

The application of AI to a specific diagnosis is commonly referred to as *narrow AI* [Beaulac and Larribe, 2017], an argument commonly referenced to justify that the radiologist job is not under threat from AI. The proponents of *narrow AI* use the example of Gamuts of radiology [Reeder, 2003], a common text used to train radiologists that references 4,600 unique imaging findings, with 13,000 unique conditions that cause findings, and 57,000 linkages between findings and conditions. When looking at the task of interpreting images, current AI papers do not describe the features of human performance. For example, radiology performance can be affected by multiple factors including availability of prior/comparison imaging, clinical history, and interruptions during actual clinical interpretations of imaging, diagnostic quality of medical imaging including artifacts, and workflow factors such as turnaround time and fatigue.

Back when McDermott at the MIT AI lab wrote about the young AI field wanting to explore weird ideas, and thus bordered between respectability and crackpottery, today that border seems past the horizon, with current researchers rarely, if ever, breaching the crackpottery border. Looking at the trend of superhuman performance claims and the debate around narrow AI, strong AI [Braga and Logan, 2017] or general intelligence [Goertzel, 2014], we ask ourselves: What wisdom does AI in medicine/radiology researchers and the tools that they build need to demonstrate? Are we repeating the same mistakes that researchers in the early days of the field made?

We propose that when human or superhuman level performance is the comparison target of training a model, then the same process used to fine tune a deep learning model to get superior results should be applied to optimizing human performance. Specifically, we explore the role of cognitive fit which postulates that when solving a problem, matching the problem representation to the task being solved results in the use of similar problem-solving processes and the formulation of a consistent mental representation, thus standardizing performance differences across multiple users. By integrating cognitive fit into medical imaging AI research, then we can determine at what points in the clinical workflow machines augment the performance of doctors to better care for patients, as well as determine the future nature of work where medical imaging experts work along machine. In fact, we argue the vice-versa too, that machine performance can be augmented by working with a human.

## 2  Philosophical expansion of AI

In this section, we do not attempt to present an exhaustive review of all the historical philosophical debate around AI, but rather build upon a tradition that mere "doing" or "understanding" or even "responding" has never been considered adequate end-goal for artificial intelligence. Haugeland (Vellino 1986) was among the first to discuss "artificial" and "natural" minds, which sparked many traditional theorists to ask what it means for a program to have a "mind of its own". Often the symbol manipulation theory [Kolers and Smythe, 1984] became the gestalt of those debates. Related to this, Searle [Searle, 2004] presented the argument through the *Chinese room* experiment that even if a computer program passed the Turing test and convinced other humans that it was able to respond back with appropriate Chinese symbols, does the machine literally understand Chinese or is it merely simulating to understand Chinese? The notion that Strong AI represents true understanding became a dominant philosophical stance at that point in time. More recently, *computationalism* [Searle, 2004] or *computer functionalism* [Harnad] is used to explain AI, where a program might be able to run program to represent mental states. We expand on this philosophy, by prescribing an expansion, that is particularly useful to the direction in which applications of AI are being developed for radiology, and medicine in many cases.

### 2.1  Phronesis and other types of wisdom

The debate around ethics of AI has just restarted, particularly with big tech leaders arguing about the risks [Mike, 2016]. Aristotle separated the wisdom as *episteme* (knowing or understanding), *techne* (craft or applied practice) and *phronesis* (prudence or discipline oneself with reason). Phronesis is thought of as "practical wisdom", that would tell a person (researcher) or their creation (AI), if they are mindful, reasonable and demonstrate moral understanding of their actions. Here, we speak of the creator and the created in the same breath because we know from sociology, and information systems that humans embed themselves in the nonhuman parts that they create [Garcia and Quek, 1997]. Technological embeddedness and more lately sociomateriality has been described using three key facets – mutuality of the human-nonhuman assemblage, performativity of when the assemblage performs and multidimensionality of where it performs [Umans, Wiskerke, and Arce 2016]. The human-nonhuman assemblage is also important from the Actor-Network theory [Müller and Schurr, 2016] (ANT) lens because it dismisses the notion that agency of an action is purely with the AI or the human. When we think of superhuman performance of AI, the assemblage is rarely explained, if ever. These studies have often separated the agency of the human (radiologist in our case), from the AI, which fails to explain when the AI should act. Thus, phronesis needs to be inculcated in the creators of AI, but also within the AI itself for it to know, when to act with the response that it has learned from reinforcement, feedforward or other kinds of learning techniques that have resulted in great strides in the last decade.

## 3  Will the radiologist be replaced by AI? The need for phronesis of AI in radiology.

The role of AI on the future of radiology is frequently reported in the media as a threat, mainly as job losses, with an unintended consequence of discouraging future trainees to explore radiology as a specialty [Lack of AI Education in Diagnostic Radiology May Be Scaring off Trainees n.d., 2018]. When authority figures in AI repeatedly make human performance and super human performance claims [arXiv n.d., 2017], they perpetuate the hype in medical AI which can result in failure of integration of AI systems in clinical care, as well as wastage of resources from building non-useful systems. One of the areas to build useful AI in medicine is to reduce diagnostic errors for medical imaging that result in high costs and death. To effectively do this, then the practice of deep learning must extend beyond computational effort to human engineering, to understand how radiologists think and practice medicine, and optimize both the human and machine for better diagnosis. We believe that the future radiologist will work with machines and will outperform the machine or human working alone. Since the cognitive errors of radiology are well studied and documented, this paper reviews the application of AI and deep learning within the context of radiology errors and biases to build a better solution where machine and human combination triumphs for future radiologists.

The cost of medical errors is high, with an estimated occurrence of between 40,000 to 400,000 events annually, costing between $17 billion and $29 billion and identified as the third leading cause of death in the United States [America et al., 2000]. Most proponents of the future of medicine specifically radiology in an era of artificial intelligence repeatedly state that the intent is not to replace doctors, but to augment their performance to reduce medical errors, save costs and save lives. Different publications describe the future of work for the radiologist in different perspectives. For example, machine learning in AI are said to be the ultimate threat to the future of radiology [J.-G. Lee et al., 2017] while other authors see a clear role to augment the performance of the radiologist [Hainc et al., 2017].

### 3.1  Where AI has failed in radiology before, we are doing much better

Computer aided diagnosis (CAD) is not new to radiology, with CAD in mammography approved for clinical use by the FDA in 1998. In a recent review of why CAD failed for breast diagnosis, low computing power and reliance on supervised learning are leading factors [Jalalian et al., 2017]. Moreover, the authors of the paper report that "CAD was trained to do what radiologists do well" – which is to pick up 84% of the breast cancers. Initially seen as a second review for radiologists, CAD compelled radiologists to look at every region flagged as abnormal. In a study reviewing 495,000 mammograms interpreted by CAD [Lehman et al., 2015], there were no improved outcomes due to the high false positives that had to be ignored by radiologists. So, for experienced radiologists, CAD generated noise and the role for the radiologist was to silence the high false positives, while for the inexperienced reader, a false sense of confidence led to misses. Given the high variation of interpretation between radiologists, its inaccurate to report human level performance or super human performance without understanding other characteristics that affect the human performance. Some of these characteristics have been described when looking at biases and misses in radiology, including workplace interruption like telephone calls, exam protocols and the general reading room environment, as well as limitations from by the type of thinking.

Current classification of deep learning studies in radiology target diagnostic performance on medical imaging data, wholly ignoring clinical context of the patient. In the CheXNet paper [Rajpurkar *et al*., 2017], the evaluated outcome is to diagnose pneumonia out of a possible 14 labels. With some clinical context, for example recent intubation or anesthesia, a radiologist would lean towards an interpretation of atelectasis, while a history of fever would favor pneumonia. It is difficult to apply such AI applications out of the box to save lives as radiology diagnosis are rarely dichotomous – e.g. normal or abnormal, but often require complex decision making. We demonstrate through our study that understanding cognitive fit in implementing deep learning in radiology can improve human and AI performance, as an assemblage, and the actual application of developed systems to the practice of health care.

### 3.2 Three "natural stupidity" that we continue to do in AI in radiology and medicine

Off course there are more, but the three we identify here are common, in the work of our own, of colleagues, and in the work of others that we have discussed in the above sections.

1. **Wishful Mnemonics**: This might be simply lack of understanding of medicine by the computer scientists or too advanced planning for the future, but we continue to see this in the reporting of AI in medicine. The CheXNet paper claimed to diagnose pneumonia, without specifying whether it is clinical or radiological pneumonia. It is common sense to understand the difference between the two type of pneumonia to any practicing clinician, as patients who are immunosuppressed are unable to mount an immune response to cause a finding on a chest x-ray. Therefore, clinical pneumonia may be negative on the radiologic pneumonia, emphasizing on the need to evaluate the whole patient picture more than a single diagnosis.
2. **Human performance as dichotomous**: While AI models are trained and improved reported as AUROC percentages, human performance is rated as dichotomous, for example in the case of the radiologist was the diagnosis pneumonia or negative. The same probabilities should be extended to human performance, adding a new level of assessment for to include certainty/confidence in a pneumonia diagnosis. While this information is not usually presented as a percentage in clinical reports, the impression section of a radiology report expresses the certainty of diagnosis by using words like *"Findings are most consistent or suggestive of malignancy, and less likely infectious etiology"*. Current literature omits this information when claiming human level or superhuman performance.
3. **Training using secondary data**: The chestxray-14 image dataset that was used for training CheXNet produced its labels using NLP from the radiology reports. This means that the NLP tool's labeling might not be correct and result in incorrect labels on which the training has been done. This is likely an example of best use of available data, but then the secondary data itself has been shown to have many problems, and thus the "superhuman" performance doesn't hold much water since it wasn't trained on accurate labels and images. The radiology report is usually used to communicate to ordering physicians, and hence an ICU frontal radiograph is commonly reported as "stable" to communicate to the ordering doctor that there is no need to adjust the support tubes and lines, but a manual review of the NLP labeling when compared to actual image findings, there is a high discrepancy of labeling. It is important to note that the practice of radiology involves looking at images, comparing with prior studies and integrating knowledge to answer a clinical question. By training models without actively looking at the images makes it difficult to translate or understand what the pixel level data generated by AI models actually means for clinical care.

### 3.3 Tackling diagnostic errors in imaging with AI and deep learning

The rapid advances in artificial intelligence (AI) and machine learning, combined with recent developments in neural network (neuromorphic) hardware technologies and ubiquitous computing, promote machines to match human-like perceptual and cognitive abilities in a way that will continue the trend of automation for several upcoming decades. Diagnostic errors occur at an average rate of 3%-4% daily, with a 32% retrospective error rate for interpretation of abnormal studies [Brady, 2016]. These high errors continue to persist despite decades of intervention. Can the integration of AI be the factor that finally results in a successful reduction of these errors? Studying the progress of AI in general where computers can learn, there is need to understand how thinking is done in radiology.

Two types of thinking are described by psychologists, fast and slow thinking or type I and II thinking. Type I or fast thinking uses only 1% of information provided, and when heuristics fail, then biases are introduced in the diagnosis. On the other hand, slow thinking is systemic, as may be seen in a trainee radiologist who will strictly adhere to a search pattern. A systematic review of the biases in radiology is summarized in Table 1. below adopted from Brusby et al, with a further postulation of how to apply cognitive fit to introduce deep learning into medicine in an applicable way.

*Table 1: Learning biases in radiology*

| Bias | Description | Cognitive fit strategies to counteract bias |
|---|---|---|
| Anchoring bias | Situation where one is fixed to their original diagnostic impression | Deep learning used to review accuracy of clinical information provided. Also used to gather more data like genomics.<br>Where high discrepancies exist with the machine diagnosis, present more cases to radiologists under "similar to" allowing for image-based search to show similar other patient cases |
| Confirmation bias | Continuous searching for data to reaffirm diagnosis. | Provide relevant and specific evidence gathered from the patient chart.<br>Present differential/alternative diagnosis to the radiologist<br>Determine what cases to present more information on e.g. not useful for a negative study |
| Availability bias | Judging probability based on what comes to mind first | Track caseloads read by a radiologist, and generate smart CMEs to help the radiologist keep up to date on the overall disease prevalence<br>Perform radiology-pathology correlation and present this in a simple way to the radiologist so they can get feedback on their study interpretations linked to outcomes. |
| Satisfaction of report | Tendency to perpetuate a diagnosis written in a prior report | Integrate deep learning techniques that can identify stable/ progression or normal/abnormal studies, and subsequently read the final report of a radiologist and determine if the change is reflected in the dictated report.<br>Develop volumetric tools for example tumor measurement that are computed on previous studies and prefilled in the radiology report with calculations such as percentage volume change.<br>Check what relevant previous studies – in and out of the hospital network that were not reviewed at a specific session and alert the radiologist if there is a significant finding that could change diagnosis. |
| Framing bias | Different diagnostic information can be drawn from the same clinical information | Help implement smart hanging protocols matching the study to be read, prior studies available and also the search pattern of the radiologist.<br>Compile clinical histories in a more relevant form and integrated to the radiology workflow. |
| Attribution bias | Based on certain stereotypes or characteristics | Process and present relevant clinical information to the radiologist |
| Satisfaction of search | Tendency to stop looking for more abnormalities once an abnormal finding is found | Learn blind spots of radiologist and act as a second check to catch errors<br>Support smart checklists and report generation |
| Premature closure | Accepting a preliminary diagnosis as final | Provide a differential diagnosis and support image search for similar cases |
| Hindsight bias | Missing findings on retrospective review | AI could provide continuous peer review by catching misses and notifying the radiologist<br>Can be used to generate teaching cases of good catches and misses |
| Inattention bias | Missing findings in plain sight due to unexpected location or nature | Learn blind spots and integrate them into a diagnostic workflow |

The two major holdups to AI creativity are said to be domain expertise and valuation of results (critical judgment of one's own idea). Looking at the concept of *experience* as machines deploy it when learning to perform tasks suggests they cannot function accurately without having humans exist side by side with them. Since the algorithms, which are the training tools of machines, learn (at least as much as the human cognitive abilities) by distilling no less than human experiences [Oussama, Norris and Amin, 2017]. Therefore, as said by many, for machines to perform well in their task, they need to be feuded with many examples of experiences, probed by humans, in order to acquire the basic ability that is required for solving the given task. Reduction in human workers will decrease the number of examples of human performance, which will then have an impact on the machine performance. It therefore becomes obvious that for machines to succeed, a continual feeding with data that reflect human experiences becomes the basic requirement. This can be achieved by integrating cognitive fit principles to make the best use of radiology-AI assemblage, that improves the performance of the assemblage as a whole.

## 4 Methodology

To demonstrate this, we developed an experiment of human-machine competition where various doctors were shown the radiographs from the chestxray-14 dataset and determine whether a chest radiograph was diagnostic or non-diagnostic of pneumonia. We were able to replicate the high variability of reporting between radiologists. We also developed additional characteristics that we are integrating into the workflow of an open source enterprise imaging system to help annotate deep learning datasets at the point of care. In this approach, we plan to integrate DICOM metadata, time, characteristics of the reader, time of study to be used as inputs for deep learning models beyond medical images. This is also enriched by collecting a snapshot of the hanging protocol and cursor movements to enable to identify areas where the radiologist is blind to form a winning man and machine combination.

As for the human-machine assemblage, the competition was between radiologists and CheXNet, the best of the breed, Chest X-ray computer vision deep learning algorithm. Radiologists with different levels of training were enrolled into the competition. CheXNet is a 121-layer convolutional neural network which can interpret probability of pneumonia from a chest X-ray image [Rajpurkar *et al.*, 2017]. It was trained on Chest X-ray 14 dataset (CXR14 here forth), which contains 112,120 anonymized frontal-view chest X-ray images.

For each image, a human or machine can select either positive or negative for pneumonia, and then select other findings like Consolidation, Infiltrates, Atelectasis etc., and a free-text area for additional comments. After submitting the interpretation, the radiologist can view their accuracy (human accuracy score) along with machine accuracy and total score. Instead of the approach where separate F-scores are reported between human and machine accuracy, we wanted to reflect a zero-sum game (like the other games where human and machines were opponents). The ground truth is something that was taken from the labels that were provided with the CXR14 dataset. Our scoring technique is based on how either the human or machine beats the other, in comparison to the ground truth:
1. If both human and machine matched with the ground truth (i.e. correct/wrong), we gave a score of 0.
2. If human is wrong and machine is correct, a score of -1 is given to the radiologist.
3. If human is correct and machine is wrong, a score of +1 is given.

A total of 607 separate competitions (image readings) were held between the human-machine assemblage. What we mean by this, is that a radiologist would not be shown the same image again, but another radiologist could be shown that image but wouldn't know the result of the previous human's interpretation. The machine on the other hand, will see the image again and learn from this new human interpretation. We also calculated the F1-scores separately for both human and machine accuracy using the Python sklearn module. Human and machine scores between the group of users' data was tested for normality using Shapiro-Wilk test, Kolmogorov-Smirnov test, Kurtosis and Skewness. We performed a two-way ANOVA to examine the impact of showing machine accuracy on different groups of humans like graduate training, years in radiology, clinical specialty, type of clinical practice on the corresponding values of human scores. All groups with 0.05 level of significance were considered significant.

## 5 Results

A total of 19 radiologists interpreted 368 unique images in 607 competitions. Four radiologists only played 1 competition and the maximum competitions played by a radiologist were 51. This radiologist had a total score of +4 after 51 competitions. Detailed scores and the number of images read by each radiologist are presented in the next table. The mean of the percentage of human and machines being accurate i.e. matching with the ground truth, on all the reports were found to be 53.26% and 50.75% respectively. Human accuracy was slightly higher but statistically insignificant. In terms of the scoring, human scores showed normal distribution, but the machine scores had left-skewness, as was expected, as the machine knew what to select, each time it lost a point, but was shown the image again. Here we can see that the AI model was getting better with each mistake and recognizing that it was not scoring higher than the human. But if the human competitor was making fewer mistakes, the machine didn't know that it was getting a zero for doing a right interpretation or a zero also when the human competitor also interpreted the image incorrectly.

*Table 2: Scores of each radiologist*

| Total score | Images read | Human accuracy (%) | Machine accuracy (%) |
| --- | --- | --- | --- |
| 8 | 43 | 53.4884 | 34.8837 |
| 7 | 51 | 47.0588 | 33.3333 |
| 7 | 19 | 73.6842 | 36.8421 |
| 5 | 11 | 54.5455 | 9.09091 |
| 4 | 51 | 60.7843 | 52.9412 |
| 3 | 15 | 60 | 40 |
| 3 | 24 | 25 | 12.5 |
| 2 | 13 | 46.1538 | 30.7692 |
| 1 | 31 | 51.6129 | 48.3871 |
| 0 | 4 | 25 | 25 |
| 0 | 5 | 20 | 20 |
| 0 | 23 | 54.1667 | 54.1667 |
| 0 | 38 | 55.2632 | 55.2632 |
| 0 | 1 | 100 | 100 |
| 0 | 9 | 55.5556 | 55.5556 |
| 0 | 1 | 100 | 100 |
| 0 | 1 | 100 | 100 |
| -1 | 1 | 0 | 100 |
| -7 | 27 | 29.6296 | 55.5556 |

Instead of a zero-sum game, using ANOVA we were able to compare the impact of machine interpretation on humans of different groups. This was done by showing the machine interpretation after each radiologist's interpretation and could be used the next time by the radiologist. This meant that now both the machine and the radiologist could work as an assemblage and help each other's scores. The machine would know why the score was given (human was right or wrong), and improve from the learning of the radiologist, and we could interpret, at least statistically, the impact that the machine's output would be on the human competitor, when the human knows the machine's accuracy (i.e. right or wrong) in advance.

Our findings showed that all the factors of graduate training, radiology years in practice or training, clinical specialty showed statistically significant ($p < 0.05$ on ANOVA) improvement (6-8%) in their scores when shown machine's accuracy on the image. Detailed appendix of all our findings can be found at arxiv preprint [arXiv n.d., 2018]. Only the time in clinical practice, for radiologists less than 5 years, it showed <1% improvement, which was not statistically significant. On the other hand, the machine accuracy improved greatly, and we could get scores of 79.5%, which is closer to what the CheXNet F1-score of 0.7820 would explain. It still did not get to full accuracy in our trail and nor did the radiologists, but it shows the potential that instead of competing, the human-machine assemblage improved performance of each other and with a large sample the possibility of improving performance could be higher.

## 6  Discussion

An obvious limitation of our study is the number of competitions that were organized. The experiment was done in an open-source, volunteer style setting, as part of a radiology in AI journal club. Yet, as the first of its kind zero-sum and then a collaborative experiment between human and machines in radiology, we learnt some interesting lessons.

When we first pitted the human versus machine as a zero-sum game, we realized that neither human or machine were getting better through the process. If it matters, we could come to a possible scenario where both human and machine are wrong, and still get a score of zero, and the machine is learning that it did the right thing, by not losing points. The problem with zero-sum games, as we know from game theories that the limiting growth factor in such competitions is the strength of the competitor.

## 7  Conclusion

AI practices in radiology and medicine specifically where imaging is involved will continuously be presented as human level or superhuman performance. However, to build AI systems that are useful in actual clinical practice, then AI in medicine needs to learn from other domains. We have demonstrated that human performance is a complex process influenced by multiple factors, causing variations in interpretation. Cognitive fit methodologies offer the AI domain a more objective and replicable framework of evaluating human performance. We describe three stupidities perpetuated across multiple studies including wishful mnemonics representing a misunderstanding of the medical domain, perception that human performance is dichotomous – yes/no or normal/abnormal and the derivation of human intent from secondary data omitting primary review of medical imaging. In this paper we have developed an assemblage that shows how to overcome the stupidities of AI and apply them to reduce cognitive errors in imaging. By incorporating phronesis in creation of AI and in AI itself, we believe that the nature of the job for the future radiologist will change, but the winning combination will always be *human and machine* vs *man or machine alone* as we describe in our human vs machine competition.

## Acknowledgments

Thanks to the radiology residents and fellows' journal club participants for being willing participants to participate in the human versus machine competition.

**APPENDIX**

**Table 1: Two-way ANOVA**

| | Two-way ANOVA | | | |
|---|---|---|---|---|
| | Sum_sq | df | F | PR(>F) |
| Doctor | | | | |
| **HA_Doctor_MD/DO** | | | | |
| MA_Doctor_MD/DO | 3314.893 | 9 | 1.88E+28 | 5.65E-15 |
| MA_Doctor_MBchB | 4.24E-28 | 3 | 0.007232 | 0.998678 |
| MA_Doctor_Not a doctor | 2.16E-27 | 5 | 0.022073 | 0.998903 |
| **HA_Doctor_MBchB** | | | | |
| MA_Doctor_MD/DO | 1.32E-27 | 9 | 0.012642 | 0.999991 |
| MA_Doctor_MBchB | 1901.001 | 3 | 5.46E+28 | 3.15E-15 |
| MA_Doctor_Not a doctor | 3.1E-27 | 5 | 0.053471 | 0.992462 |
| **HA_Doctor_Not a doctor** | | | | |
| MA_Doctor_MD/DO | 3.79E-27 | 9 | 0.019251 | 0.99995 |
| MA_Doctor_MBchB | 1.19E-27 | 3 | 0.018183 | 0.994927 |
| MA_Doctor_Not a doctor | 2239.321 | 5 | 2.05E+28 | 5.3E-15 |
| Radiologist | | | | |
| **HA_Radiologist_Yes** | | | | |
| MA_Radiologist_Yes | 3269.691 | 10 | 5.9E+28 | 3.68E-86 |
| MA_Radiologist_No | 3.11E-28 | 2 | 0.028036 | 0.97248 |
| **HA_Radiologist_No** | | | | |
| MA_Radiologist_Yes | 9.35E-27 | 10 | 0.224062 | 0.981287 |
| MA_Radiologist_No | 4266.667 | 2 | 5.11E+29 | 2.02E-88 |
| **Clinical Specialty** | | | | |
| **HA_Clinical Speciality_Body/Abdomen** | | | | |
| MA_Clinical Speciality_Body/Abdomen | 26.61508 | 2 | 1.09E+29 | 4.6E-198 |
| MA_Clinical Speciality_ER General | 7.98E-28 | 2 | 3.253438 | 0.069119 |
| **HA_Clinical Speciality_ER General** | | | | |
| MA_Clinical Speciality_Body/Abdomen | 1.44E-27 | 2 | 0.682504 | 0.521395 |
| MA_Clinical Speciality_ER General | 3200 | 2 | 1.52E+30 | 4.4E-206 |
| **Training** | | | | |
| **HA_Training_r1** | | | | |
| MA_Training_r1 | 1065.087 | 2 | 2.97E+29 | 5.6E-102 |
| MA_Training_r3 | 9.08E-28 | 4 | 0.1266 | 0.968038 |
| MA_Training_staff | 1.35E-27 | 5 | 0.150145 | 0.973458 |
| **HA_Training_r3** | | | | |
| MA_Training_r1 | 7.63E-28 | 2 | 0.332503 | 0.727863 |
| MA_Training_r3 | 3820.862 | 4 | 8.33E+29 | 6E-104 |
| MA_Training_staff | 1.75E-27 | 5 | 0.304712 | 0.895317 |
| **HA_Training_staff** | | | | |
| MA_Training_r1 | 1.87E-27 | 2 | 0.325525 | 0.73252 |
| MA_Training_r3 | 4.81E-27 | 4 | 0.419109 | 0.79066 |
| MA_Training_staff | 48.08411 | 5 | 3.35E+27 | 1.16E-95 |
| **Clinical Practice** | | | | |
| **HA_Clinical practice_<5 years** | | | | |
| MA_Clinical practice_<5 years | 1099.19 | 6 | 0.256478 | 0.94104 |
| MA_Clinical practice_5 to 10 years | 1.79E-27 | 5 | 5E-31 | 1 |
| **HA_Clinical practice_5 to 10 years** | | | | |
| MA_Clinical practice_<5 years | 1.03E-27 | 6 | 0.24705 | 0.945616 |
| MA_Clinical practice_5 to 10 years | 1012.422 | 5 | 2.92E+29 | 1.9E-102 |

**Table 2: One-Way ANOVA:**

| One-way ANOVA | | | |
|---|---|---|---|
| | Df | F | Pr(>F) |
| **Human_accuracy** | | | |
| Doctor | 3.0 | 3.90 | 0.03 |
| Radiologist | 1.0 | 3.83 | 0.07 |
| Training | 4.0 | 0.93 | 0.49 |
| Clinical_practice | 1.0 | 0.16 | 0.69 |
| Clinical_speciality | 5.0 | 0.41 | 0.81 |
| Institution_type | 1.0 | 0.57 | 0.46 |
| Country | 5.0 | 1.17 | 0.37 |
| State | 6.0 | 0.63 | 0.69 |
| **Machine_accuracy** | | | |
| Doctor | 3.0 | 0.98 | 0.42 |
| Radiologist | 1.0 | 2.53 | 0.13 |
| Training | 4.0 | 0.69 | 0.61 |
| Clinical_practice | 1.0 | 0.36 | 0.55 |
| Clinical_speciality | 5.0 | 0.60 | 0.71 |
| Institution_type | 1.0 | 0.19 | 0.67 |
| Country | 5.0 | 1.08 | 0.41 |
| State | 6.0 | 0.82 | 0.57 |

**Table 3: Normality testing**

| | Kurtosis | Skewness | Shapiro-Wilk test | Kolmogorov-Smirnov test |
|---|---|---|---|---|
| Human_accuracy | -0.265 | 0.196 | 0.929, 0.167 | 0.947, 0.0 |
| Machine_accuracy | -0.705 | 0.616 | 0.880, 0.021 | 1.0, 0.0 |

**Table 4: F1-Scores:**

| | F1-Score |
|---|---|
| Human_accuracy | 0.5714285714285714 |
| Machine_accuracy | 0.59653179190751437 |